\documentclass[]{aa}

\usepackage{graphicx}
\usepackage{txfonts}
\usepackage{multirow}

\begin{document}

   \title{A search for supernova-like optical counterparts to ASKAP-localised Fast Radio Bursts}


   \author{Lachlan~Marnoch
          \inst{1,2}
          \and
          Stuart~D.~Ryder\inst{1}
          \and 
          Keith~W.~Bannister\inst{2}
          \and
          Shivani~Bhandari\inst{2}
          \and
          Cherie~K.~Day\inst{3,2}
          \and
          Adam~T.~Deller\inst{3}
          \and
          Jean-Pierre~Macquart\inst{4}
          \and
          Richard~M.~McDermid\inst{1,5}
          \and
          J.~Xavier Prochaska\inst{6,7}
          \and
          Hao Qiu\inst{8,2}
          \and
          Elaine~M.~Sadler\inst{8,2}
          \and
          Ryan~M.~Shannon\inst{3}
          \and
          Nicolas~Tejos\inst{9}
          }

   \institute{Department of Physics and Astronomy, Macquarie University, NSW 2109, Australia\\
              \email{lachlan.marnoch@hdr.mq.edu.au}
         \and
             Australia Telescope National Facility, CSIRO Astronomy and Space Science, P.O. Box 76, Epping, NSW 1710, Australia
        \and
             Centre for Astrophysics and Supercomputing, Swinburne University of Technology, Hawthorn, VIC 3122, Australia
        \and
             International Centre for Radio Astronomy Research, Curtin Institute of Radio Astronomy, Curtin University, Perth, WA 6845, Australia
        \and
             Astronomy, Astrophysics and Astrophotonics Research Centre, Macquarie University, Sydney, NSW 2109, Australia
        \and
             University of California Observatories--Lick Observatory, University of California, Santa Cruz, CA 95064, USA
        \and
             Kavli Institute for the Physics and Mathematics of the Universe, 5-1-5 Kashiwanoha, Kashiwa 277-8583, Japan
        \and
             Sydney Institute for Astronomy, School of Physics, University of Sydney, NSW 2006, Australia
        \and
             Instituto de F\'isica, Pontificia Universidad Cat\'olica de Valpara\'iso, Casilla 4059, Valpara\'iso, Chile
             }

   \date{Received 03 April 2020; accepted 03 June 2020}

 
  \abstract{
   Fast radio bursts (FRBs) are millisecond-scale radio pulses, which originate in distant galaxies and are produced by unknown sources. 
The mystery remains partially because of the typical difficulty in localising FRBs to host galaxies.
Accurate localisations delivered by the Commensal Real-time ASKAP Fast Transients (CRAFT) survey now provide an opportunity to study the host galaxies and potential transient counterparts of FRBs at a large range of wavelengths.
   In this work, we investigate whether the first three FRBs accurately localised by CRAFT have supernova-like transient counterparts.
   We obtained two sets of imaging epochs with the Very Large Telescope for three host galaxies, one soon after the burst detection and one several months later. After subtracting these images no optical counterparts were identified in the associated FRB host galaxies, so we instead place limits on the brightness of any potential optical transients. A Monte Carlo approach, in which supernova light curves were modelled and their base properties randomised, was used to estimate the probability of a supernova associated with each FRB going undetected. We conclude that Type Ia and IIn supernovae are unlikely to accompany every apparently non-repeating FRB.}


   \maketitle
%

    
\section{Introduction}

   Fast radio bursts (FRBs) are bright, millisecond-timescale
radio frequency pulses of extragalactic origin \citep{Keane2018, Chatterjee2017, FRB180924}. With a detectable all-sky rate on the order of a thousand per day \citep{Bhandari2018}, these events are clearly common in the universe -- and yet we remain in the dark as to what generates them. Recent sub-arcsecond localisations of apparently non-repeating FRBs \citep{FRB180924, FRB181112, HostGalaxies, CosmicDM} by the Commensal Real-time ASKAP Fast Transients (CRAFT; \citep{CRAFT}) survey finally allow us to study the host galaxies of FRBs throughout the entire electromagnetic range.

No non-radio transient counterpart to an FRB has yet been detected \citep{Petroff2019}. 
Despite this, several FRB hypotheses call for association with some kind of luminous transient event, such as
supernovae (SNe; \citealp{Kashiyama2013}) or
kilonovae \citep{Wang2016, Totani2013, Yamasaki2018}. Although repeating sources exist \citep{CHIME2019a, CHIME2019b, Spitler2016}, the majority of sources have not been found to repeat; there is also evidence to suggest that there are at least two distinct populations of FRBs \citep{CHIME2019a}. Non-repeating bursts could, potentially, be associated with cataclysmic events.
Indeed, given the phenomenal energy implied by the brightness of the bursts, cataclysmic events such as these are especially appealing for apparent non-repeaters. 
Candidate models include magnetic reconnection due to mergers of white dwarfs \citep{Kashiyama2013} or neutron stars \citep{Wang2016, Totani2013} and interactions between supernova shocks and neutron stars \citep{Egorov2009}. 

Event rates offer some circumstantial evidence for a link to such events.
Some analyses suggest that the FRB rate is consistent with the merger rate of binary neutron stars, which could produce kilonova counterparts \citep{Totani2013, Abbott2017, Petroff2019}. \citet{Kashiyama2013} suggest it is consistent with the merger rate of binary white dwarfs, which could produce Type Ia SN counterparts.
Based on the inferred volumetric distribution of FRBs, the event rate of Type Ib/c SNe is consistent with that of FRBs to within an order of magnitude \citep{Dahlen2012, Petroff2019, Gupta2018}, and the rate of core-collapse SNe in general is higher by two orders of magnitude \citep{Dahlen2004, Petroff2019}.
Depending on the extent to which FRBs are beamed or isotropic, each of these events offers a viable FRB counterpart. 

Previous searches for optical FRB counterparts have been undertaken; however, all were on bursts with uncertain localisation. A study of the field containing FRB\,151230 was undertaken to locate optical transient counterpart candidates \citep{Tominaga2018}. 
It found no less than 8 candidates in the search volume, which \citet{Tominaga2018} concede may all be unrelated to the burst. 
The study also ruled out association of that particular burst with a Type Ia supernova to $z\leq0.6$, although the high dispersion measure (DM) of $960~\mathrm{pc}\ \mathrm{{cm}}^{-3}$ indicates that the host redshift is likely to be greater than this value \citep{Tominaga2018, Inoue2004}.

The field of FRB\,140514 was checked thoroughly for transients in X-ray, near-infrared, optical and radio bands, ruling out an accompanying supernova or long GRB to redshift 0.3 \citep{Petroff2015a}.  This does not, however, rule out a supernova at greater distances, and the DM of this burst (562 $\mathrm{pc}\ \mathrm{{cm}}^{-3}$) is consistent with a redshift as high as 0.5 \citep{Petroff2015a, Inoue2004}. The observations took place from 8.5 hours out to 55 days after the burst \citep{Petroff2015a}.

The main limitation of previous follow-up searches for optical FRB counterparts has been, due to poor localisation, the lack of a definitively associated host galaxy. The growing sample of accurately-localised FRBs now allows a more precise search, relying on optical follow-up performed in the weeks following the burst. With such data, constraints can be placed on the brightness of potential optical transients directly from observations of a known host galaxy. In this letter, we target the hosts of three FRBs, detected by the Australian Square Kilometre Array Pathfinder (ASKAP) and localised by CRAFT: FRB\,180924 \citep{FRB180924}, FRB\,181112 \citep{FRB181112} and FRB\,190102 \citep{CosmicDM, HostGalaxies}. 

These were each localised from a single burst detection, and have not been found to repeat despite extensive follow-up in the same fields \citep{James2019}. Although it is possible that one or more of them emit repeat bursts, we note that all 3 had pulse widths somewhat smaller than found in the known repeaters \citep{CHIME2019b}, as well as differing in their time-frequency structure and polarization properties (Day et al. 2020; \citealp{Cho2020}). 
\textbf{We caution that the conclusions to be drawn here apply specifically to the majority population of non-repeating (or only rarely repeating) FRBs.}

\section{Observations and Results}
%
%
%

\begin{table}
\begin{center}
\caption{CRAFT-localised FRB detection and subsequent VLT observation dates. Full details of the FRBs can be found in \citet{FRB180924, FRB181112, CosmicDM}; and \citet{HostGalaxies}}
\label{tab:comparison_observations}
\begin{tabular}{lcccl}
\hline\hline
FRB     & Burst detection   & Observation  & $\Delta t$  \\
        & (MJD)             & (MJD)        & (days) 
\\ \hline\hline
\multirow{2}{*}{180924} & \multirow{2}{*}{58385}    & 58431     & 46     
\\                      &                           & 58718     & 333   
\\ \hline
\multirow{2}{*}{181112} & \multirow{2}{*}{58434}    & 58455     & 20    
\\                      &                           & 58718     & 284
\\ \hline
\multirow{2}{*}{190102} & \multirow{2}{*}{58485}    & 58495     & 10
\\                      &                           & 58718     & 233
\\ \hline\hline
\end{tabular}
\end{center}
\end{table}

\begin{figure*}
    \centering
         \includegraphics[width=0.7\textwidth]{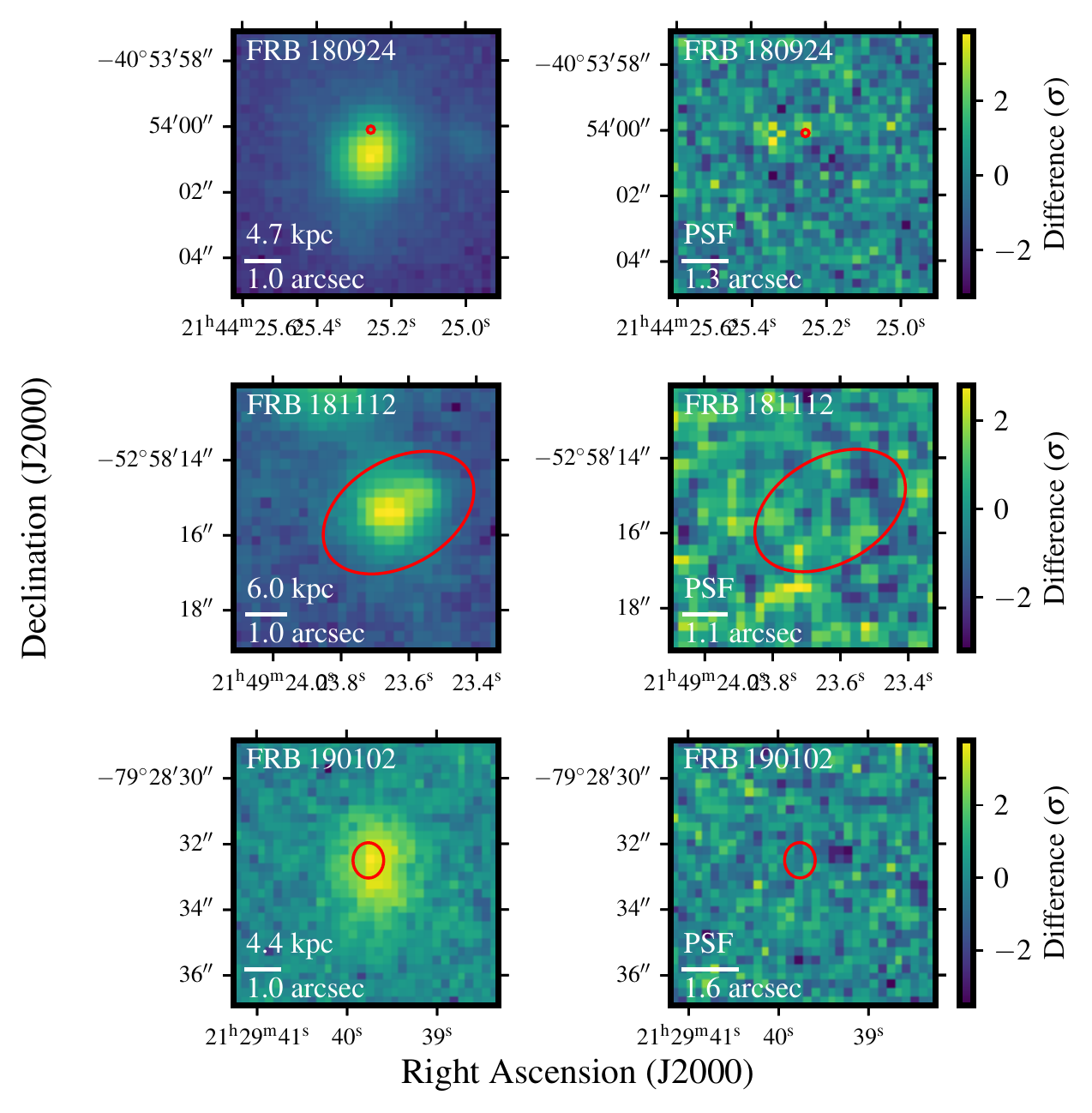} 
         \caption{VLT/FORS2 $g$-band imaging of the host galaxies for (from top to bottom) FRB\,180924, FRB\,181112 and FRB\,190102. The left-hand column is the first epoch image, and the right-hand column is the difference image resulting from the subtraction of the second epoch from the first. Red ellipses show the uncertainty on the FRB position in each case. Colour bars show the pixel values of the difference image, in units of standard deviation from the mean. The bars in the lower-left corner of each image show the image scale. In the left column, they are fixed at 1 arcsec; in the right column they are set to the greater PSF FWHM of the two epochs used in subtraction, which corresponds to the effective PSF of the difference image.}
         \label{fig:subtractions_g}
\end{figure*}



\subsection{Data}

The imaging data were taken on the Focal Reducer and low dispersion Spectrograph 2 (FORS2; \citealp{FORS}), mounted on Unit Telescope 1 (UT1) of the Very Large Telescope (VLT). 
Two observation epochs were taken for each host galaxy:  the first within a few weeks of burst detection, near the time at which a SN-like event would be expected to peak optically; and another a few months later, when any SN-like event would have faded below the detection limit.
These are summarised in Table \ref{tab:comparison_observations}, while the full program of CRAFT follow-up observations to date is given by \citet{HostGalaxies}. 
For each observation, $5 \times 500 \mathrm{s}$ exposures were obtained in $g$-band, except for the first FRB 190102 epoch, for which only three exposures were taken; and $5 \times 90$ s exposures were obtained in $I$-band for all hosts. 
These were reduced and co-added using the 
process\footnote{The code written for this pipeline, as well as for the research described later in this letter (including the software parameter files used, such as for SExtractor), is available at 
https://github.com/Lachimax/craft-optical-followup}
described in \citealp{FRB181112}, \citealp{CosmicDM} and \citealp{HostGalaxies}, using the packages ESO Reflex \citep{ESOReflex}, Montage \citep{Montage}, {\sc astropy} \citep{astropy1} and Astrometry.net \citep{Astrometry}. 

\subsection{Image Subtraction} \label{sec:subtraction}

To search our data for optical transients, we subtracted the second image from the first using the package {\sc hotpants} \citep{Hotpants}, which implements the \citealp{Alard2000} algorithm for PSF-matching and image subtraction. The difference images in $g$-band produced by this process are shown in Figure \ref{fig:subtractions_g}.
The difference image was then searched for sources using SExtractor \citep{SExtractor}.
Although residuals do appear in the difference images, they are narrower than the point-spread functions of the respective images (as marked in Figure \ref{fig:subtractions_g}) and are thus likely to be subtraction artefacts.
There are no residuals that are within the localisation ellipse, span the PSF of the difference image, and are significant to greater than $3\sigma$. We can thus make no positive identification of a transient optical counterpart to any of the three FRBs.

\begin{figure*}
   \centering
   \includegraphics{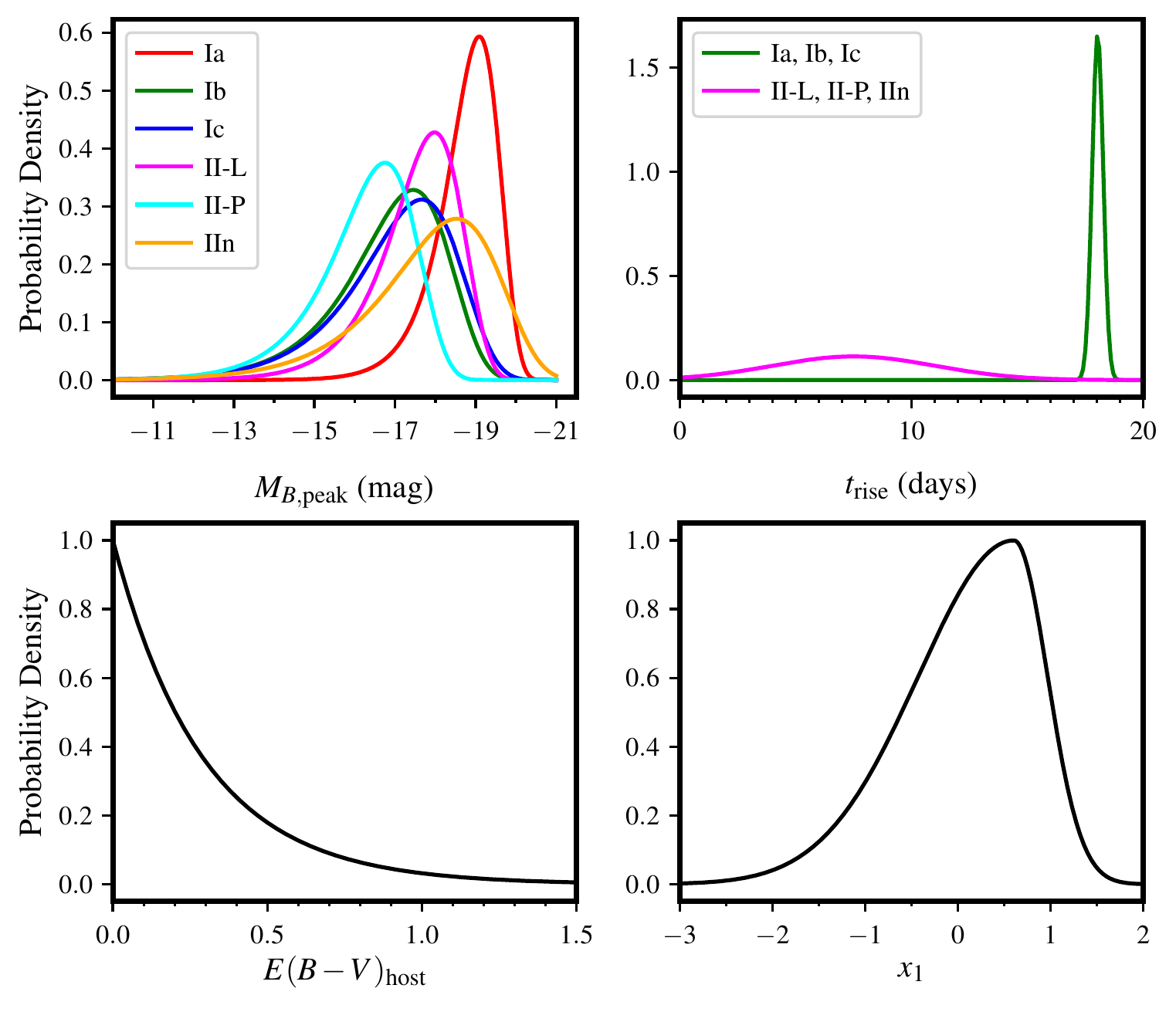}
   \caption{Curves of the probability density functions from which the various parameters used for the Monte Carlo analysis were drawn. Clockwise, starting at the top-left, these are: the peak magnitude of the supernova in \textit{B}-band, $M_{B,\mathrm{peak}}$; the rest-frame rise time of the supernova, $t_\mathrm{rise}$; the Type Ia SN stretch parameter $x_1$; and the host galaxy reddening parameter $E(B-V)_\mathrm{host}$. }
    \label{fig:distributions}
\end{figure*}

\section{Analysis}

\subsection{Sensitivity testing}

Synthetic point sources were used to test the sensitivity of our method to supernova-like transients. 
The point-spread functions of the images were modelled using PSFEx \citep{PSFEx} and used to insert a synthetic point-source with the desired flux and at the desired position in the first image, before subtraction. After subtraction as normal, the difference image was searched blindly using SExtractor to test our ability to recover the synthetic source. 

For a coarse upper brightness limit, sources with a range of apparent magnitudes were inserted at the FRB position using this method. The magnitude of the faintest source detected by SExtractor after subtraction was taken as the upper brightness limit at the burst position, as given in Table~\ref{tab:results_limits} (typically $g \gtrapprox 25$). 

\subsection{Monte Carlo experiment} \label{sec:monte-carlo}

\begin{figure*}
    \centering
    \includegraphics{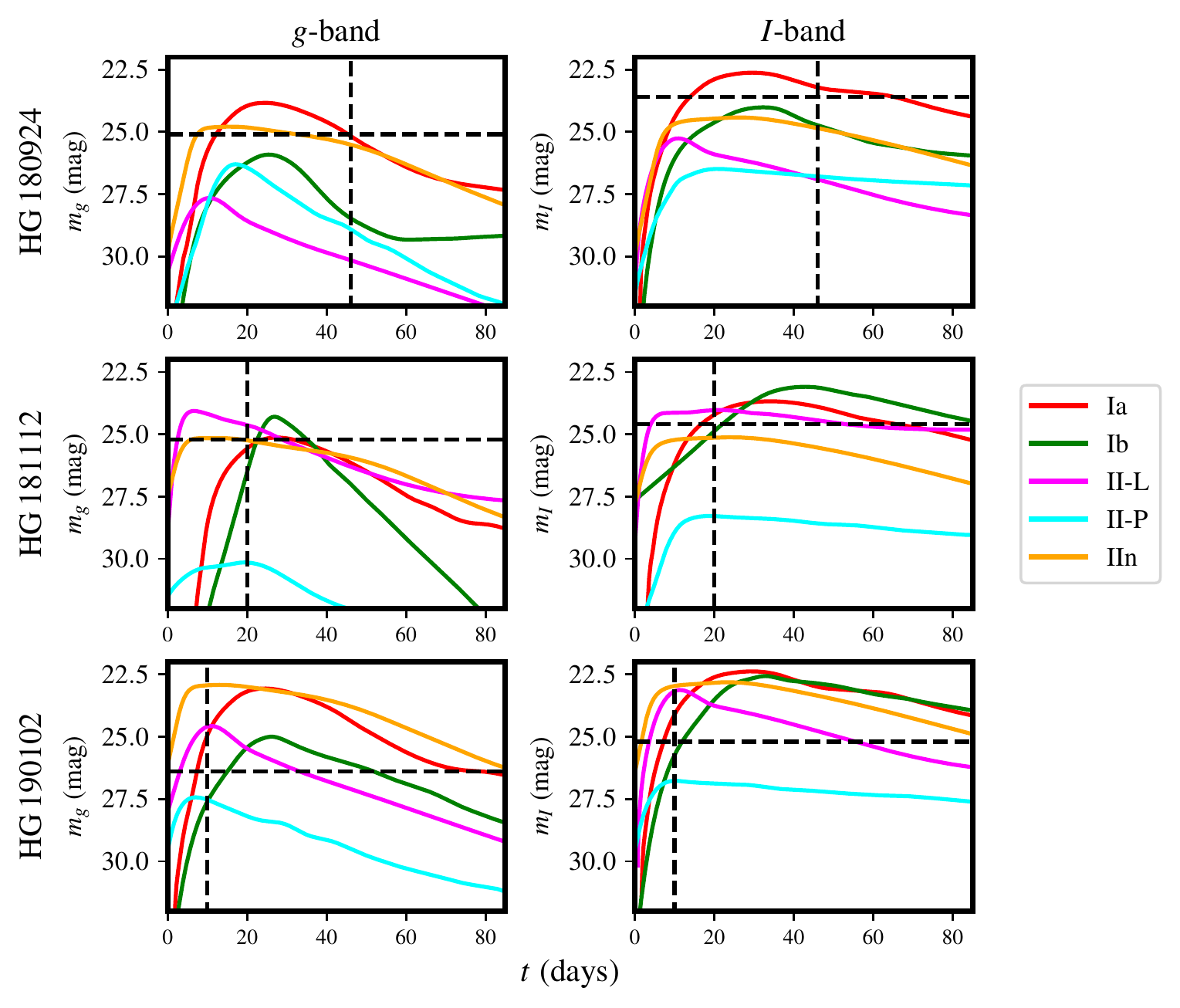}
    \caption{A selection of randomised, observer-frame supernova light curve models (apparent magnitude) from the Monte Carlo analysis sample. Both Milky Way reddening along the line-of-sight and randomised host-galaxy reddening were included in the generation of these curves. The vertical dashed lines indicate the epoch of the first VLT observation of each source, and the horizontal dashed line is the detection limit reported in Table \ref{tab:results_limits}.}
    \label{fig:light_curves}
\end{figure*}

The limits derived using the above method are only truly valid when the source position is well-localised within the host galaxy; that is, only for FRB\,180924 \citep{FRB180924}. The localisation ellipses of the other two FRBs are significantly larger \citep{FRB181112, CosmicDM}, and the uncertainty in the position of FRB\,181112, in particular, could place it almost anywhere within its host galaxy. A more robust measure of the likelihood of detection, with some assumptions about the nature of the optical counterparts themselves, was hence required.

A Monte Carlo analysis was conducted to estimate the probability of detecting a supernova in the difference imaging. The Python package SNCosmo \citep{sncosmo} was used to generate model light curves for SN types Ia, Ib, Ic, II-L, II-P and IIn, and to move them to the host redshift. To replicate the variety in the properties of observed supernovae, other parameters were drawn probabilistically from distributions based on the astrophysical literature. These are: the position of the explosion within the host galaxy; the luminosity of the supernova via the peak absolute magnitude, $M_{B,\mathrm{peak}}$; the rise-time of the supernova in its rest frame, $t_\mathrm{rise}$; the reddening of the supernova due to host galaxy dust, $E(B-V)_\mathrm{host}$; for Type Ia SNe, the `stretch' of the light-curve via the SALT2 parameter $x_1$;
and, for all SN types besides Ia, the fundamental shape of the light curve. The distributions from which these parameters were selected are visualised in Figure \ref{fig:distributions} and described below. 

Galactic reddening was also applied by SNCosmo; we set $E(B-V)_\mathrm{MW}$ for each line-of-sight according to the IRSA Dust Tool\footnote{https://irsa.ipac.caltech.edu/applications/DUST/}
from the \citet{SandF} reddening map, which applies the \citet{F99} extinction law.

We assumed that the FRB was emitted contemporaneously with the putative supernova explosion. A synthetic source with a brightness set by the appropriate point in time (corresponding to the first FORS2 epoch) on the generated light curve was inserted in the first image. The second epoch image was subtracted, consistent with our original analysis. If SExtractor was able to recover the coordinates (to within 1 arcsecond) of the inserted synthetic source from the difference image, the trial was counted as a detection.

100 trials were performed for each host galaxy and for each SN type supported by SNCosmo, with the parameters of each trial drawn independently. Each trial SN was inserted and tested for detectability in both $g$ and $I$-band. The percentage of successes was taken as the probability of detection. Conversely, the percentage of failures (i.e. non-detections) is reported (Table~\ref{tab:results_limits}) as the probability that, if a SN of the given type had occurred in the host galaxy at the epoch of FRB emission, we would have failed to detect it in our imaging. 

\subsection{Model parameters}

\subsubsection{Light curve profile}

As Type Ia profiles are relatively uniform, the implementation of the SALT2 model \citep{SALT} supplied with SNCosmo was used for all Type Ia insertions.
For other SN types, to capture the diversity of observed light curve profiles for core-collapse supernovae, the particular SNCosmo source (and hence the particular light-curve shape) was selected at random from those available for the given 
type\footnote{All built-in models are listed here: \\ https://sncosmo.readthedocs.io/en/v2.0.x/source-list.html}. 

\subsubsection{Type Ia stretch $x_1$}

The `stretch' parameter of a Type Ia supernova -- that is, the rate of decay in the event's apparent magnitude after peak -- is captured in the dimensionless $x_1$ parameter in SALT2 \citep{SALT, HandbookTypeIa}. As this influences the supernova's detectability as a function of time, this parameter was varied as well. 
The distribution model used by \citet{Scolnic2016}, dubbed an `asymmetric Gaussian', was adopted here. 
We used the values given for the SNLS survey \citep{SNLS}, the median redshift $z=0.29$ \citep{Scolnic2016} of which is the nearest of their samples to our FRB host galaxy values. The distribution obeys $\sigma_+ = 0.363$, $\sigma_-=1.029$, and $\mu=0.604$. 

\subsubsection{Peak absolute magnitude $M_{B,\mathrm{peak}}$}

The peak absolute magnitude was set in $B$-band; SNCosmo then uses this to set the brightness of the light curves in other band-passes (in this case, FORS2 $g$ and $I$) according to the source model. 
Here, we used a right-handed Gumbel 
distribution in an attempt to capture the empirical, subluminous tail of peak magnitude distributions \citep{Ashall2016, Richardson2014}. The Type Ia distribution is based on that found by \citet{Ashall2016}, with $\mu(M_{B,\mathrm{peak}}) =-19.09$ and $\sigma(M_{B,\mathrm{peak}})=0.62$. Other values come from the bias-corrected sample of \citet{Richardson2014}.

\subsubsection{Rest-frame rise time $t_\mathrm{rise}$}

The rise times $t_\mathrm{rise}$ of Type Ia, Ib and Ic SNe are thought to be similar \citep{Richardson2002}, so the same distribution was adopted for all three. Assuming a normal distribution, we adopted $\mu(t_\mathrm{rise}) = 18.03$ days with $\sigma(t_\mathrm{rise})=0.24$ days from \citet{Ganeshalingam2011}. For Type II SNe, using values quoted in \citet{Gonzalez-Gaitan2015}, we took $\mu(t_\mathrm{rise}) = 7.5$ days, and, in an attempt to capture the wide range in observed Type II rise times, a standard deviation of 3.5 days. 

To correct coarsely to the selected rest-frame 
rise-time,
the light curves were simply shifted in time so that $B$-band peak luminosity occurs, in the observer frame, at $t_\mathrm{peak} = (1+z)\times t_\mathrm{rise}$ \citep{Zhang2013}.

\begin{table*}
    \centering
    \caption{Results of synthetic insertion tests. The probability of non-detection is the estimated probability that, if a supernova of this type did occur in the given host galaxy coincident with the FRB emission, it would not be found. The \textit{combined} column gives the estimated probability that, if a given SN type occurred in all three difference images, it would go undetected.}
    \label{tab:results_limits}
    \begin{tabular}{l|cc|cc|cc|cc}
     FRB            & \multicolumn{2}{c|}{180924}    & \multicolumn{2}{c|}{181112} & \multicolumn{2}{c|}{190102}  & \multicolumn{2}{c}{Combined}
     \\ \hline
     Band               & $g$  & $I$  & $g$  & $I$  & $g$  & $I$  & $g$  & $I$ 
     \\ \hline
     Limit at burst position (mag) & 25.1 & 23.6 & 25.2 & 24.6 & 26.4 & 25.2 & - & -
     
     \\ \hline
     \multicolumn{9}{c}{Probability of non-detection}
     \\ \hline
     Type Ia            & 34\% & 12\% & 44\% & 34\% & 78\% & 88\% & 12\% & 4\%  
     \\
     Type Ib            & 94\% & 54\% & 91\% & 91\% & 99\% & 99\% & 85\% & 49\% 
     \\
     Type Ic            & 94\% & 49\% & 88\% & 80\% & 98\% & 98\% & 82\% & 39\%
     \\
     Type IIn           & 40\% & 41\% & 34\% & 47\% & 39\% & 65\% & 6\%  & 13\%
     \\
     Type II-L          & 69\% & 48\% & 78\% & 79\% & 63\% & 76\% & 34\% & 29\% 
     \\
     Type II-P          & 89\% & 73\% & 75\% & 89\% & 74\% & 90\% & 50\% & 59\%
     
     \\ \hline
     
\end{tabular}
\end{table*}

\subsubsection{Reddening due to host galaxy dust $E(B-V)_\mathrm{host}$}
\label{redenning}

For reddening of the supernova due to host galaxy dust, we used the \citet{CCM89} extinction law, adopting the extinction distribution given by \citet{Holwerda2014}. This takes the form 
$N = N_0 \exp\left(\frac{-A(V)}{\tau}\right)$, where $N$ is the number density of supernovae with extinction $A(V)$.
With no a priori knowledge of the inclination of our host 
galaxies,
we took the value of $\tau = 0.67$, which corresponds to the distribution quoted by \citet{Holwerda2014} prior to correction for inclination. Using the relationship $R_V = \frac{A(V)}{E(B-V)}$, and setting $R_V = 2.3$ as given by \citet{Holwerda2014} for supernovae, we used the distribution of $N = N_0 \exp\left(\frac{-2.3E(B-V)}{0.67}\right)$ to select $E(B-V)_\mathrm{host}$ along the supernova line-of-sight. 

In order to check the validity of the range of host galaxy extinction used here, an estimate can be made using the host galaxy's contribution to the burst DM, $\mathrm{DM}_\mathrm{host}$. Although this value is hard to constrain for any single FRB, \citet{Li2020} estimate an average of $34^{+39}_{-31}\ \mathrm{pc}\ \mathrm{{cm}}^{-3}$ for three localised FRBs, including FRB\,180924 and FRB\,181112. Using this value, the $N_\mathrm{H}$--$A_V$ relation given by \citet{Guver2009}, and the $N_\mathrm{H}$--DM relationship found by \citet{He2013}, we estimate $E(B-V)_{\mathrm{host}} \approx 0.2^{+0.4}_{-0.2}$. This estimate is consistent with the distribution of values used in our experiment.

\subsubsection{Position}

The position of the synthetic transient was selected with the simplifying assumption that the rate of supernovae at a given position within a galaxy is proportional to the stellar density in that region; and that, in turn, the stellar density is proportional to the $g$-band surface brightness at that position. Although certainly sufficient for Type Ia SNe, which trace the $B$-band light distribution of their hosts \citep{Anderson2015}, the first assumption does not hold entirely for core-collapse SNe, which tend to trace star formation rather than the general stellar population \citep{Anderson2009}.
However, in lieu of a high-resolution star formation map, the surface brightness in $g$-band (which traces the younger stellar population better than the $I$-band) served as an adequate proxy. The position was then chosen from a distribution defined by the image of the galaxy itself; each pixel in $g$ was assigned a probability, proportional to its flux, of hosting the supernova. The same position was then used to insert the source into both $g$ and $I$ images.

\subsection{Results and discussion}

Results of our Monte Carlo analysis are summarised in Table~\ref{tab:results_limits}.
For FRB\,180924, there is an 88\% probability that we would have detected a Type Ia supernova had it taken place, based on the strongest constraint (in $I$-band). The estimate is less stringent for FRB\,181112, but still favours detection at 66\%. The results for Types Ib and Ic are not very constraining for either FRB\,180924 or 181112. For Type IIn SNe, the results mildly favour detection for each host and disfavour it for Type II-L and II-P.

Our initial observations of FRB\,190102 took place 10 days after the burst detection, several days before the Type Ia, Ib or Ic peak epoch in all cases.
This is the reason for the low detection probability for all Type I SNe in the host galaxy of FRB\,190102. 
It is hence unlikely that any Type I SN would have been detected in the imaging of this host. However, the fast mean rise time taken from  \citet{Gonzalez-Gaitan2015} often places the first 190102 epoch quite close to peak brightness for Type II-L, II-P and IIn SNe. Hence, the probability of detection rises significantly for Type II in the 190102 imaging. Similar to FRBs 180924 and 181112, detection remains unlikely for Types II-L and II-P, but a Type IIn is likely to have been detected in $g$.

We find it unlikely that FRBs 180924 or 181112 occurred contemporaneously with Type Ia SNe. We cannot make a similar conclusion about FRB\,190102. It is also unlikely that a Type IIn SN accompanied any of the three bursts. The results for Type Ib/c and II-L/P are less limiting in all individual cases. 

Although we cannot rule out that any \textit{one} of FRBs 180924, 181112 or 190102 occurred simultaneously with a supernova in the associated host galaxies, when they are considered together, it becomes increasingly unlikely that all three did so. There is only a 4\% probability that a Type Ia supernova would have gone undetected in all three $I$-band difference images; the figure for Type IIn in $g$-band is 6\%. Type II-L SNe are also less likely when considered like this, with an $I$-band non-detection probability of 29\%. So, although any one of the FRBs considered here could have had a SN-like counterpart, it is quite unlikely that all of them occurred contemporaneously with Type Ia or Type IIn SNe. It is also improbable that all of them occurred simultaneously with Type II-L SNe.

\section{Conclusions} \label{sec:results}

The primary results of this study are:

   \begin{enumerate}
      \item No positive detection of an optical transient counterpart to a set of three apparently non-repeating FRBs was found.
      \item Although not definitively ruled out, we find that it is unlikely that every apparently non-repeating fast radio burst is coincident with a Type Ia or Type IIn supernova explosion, or with another type of slow optical transient with a similar light curve.
   \end{enumerate}

Despite these findings, this work cannot make any statement about whether these FRBs are caused by supernova shocks at later, fainter, epochs; or whether FRBs significantly precede SN outbursts. We also cannot eliminate kilonovae as optical counterparts to FRBs, as their faintness would put them well below the detection thresholds reported here. 
Although we did not consider superluminous supernovae owing to the diversity of their behaviours, and can make no definitive statement about them, the fact that these events are considerably more luminous than those SNe which we did model, and remain so for longer \citep{Gal-Yam2012, Gal-Yam2019}, makes it seem unlikely that these FRBs had contemporaneous superluminous counterparts.

We cannot rule out that FRBs are associated with novel optical transients that are fainter than the derived limits, too brief to appear in our imaging, or both. Deeper imaging is required to investigate the first possibility and prompt optical follow-up for the second. Now that we are entering an age in which the study of known FRB host galaxies is possible, these investigations can be performed.

\begin{acknowledgements}

We thank the referee Victoria Kaspi for suggestions that improved this paper. 
Based on observations collected at the European Southern Observatory under ESO programme 0102.A-0450(A).
The Australian Square Kilometre Array Pathfinder and Australia Telescope Compact Array are part of the Australia Telescope National Facility which is managed by CSIRO. 
Operation of ASKAP is funded by the Australian Government with support from the National Collaborative Research Infrastructure Strategy. ASKAP uses the resources of the Pawsey Supercomputing Centre. Establishment of ASKAP, the Murchison Radio-astronomy Observatory and the Pawsey Supercomputing Centre are initiatives of the Australian Government, with support from the Government of Western Australia and the Science and Industry Endowment Fund. 
L.M. acknowledges the receipt of an MQ-MRES scholarship from Macquarie University.
H.Q. is supported by the Hunstead Merit Award in Astrophysics from the University of Sydney.
K.W.B., J.P.M, and R.M.S. acknowledge Australian Research Council (ARC) grant DP180100857.
A.T.D. and R.M. are recipients of ARC Future Fellowships (FT150100415 \& FT150100333, respectively).
J.X.P. is supported by NSF AST-1911140, while N.T. acknowledges support from FONDECYT grant number 11191217.

\end{acknowledgements}

%
%

\bibliographystyle{aa}
\bibliography{references_new}






   
  



\end{document}